\documentclass[sensors,article,submit,moreauthors,pdftex]{mdpi} 

\firstpage{1} 
\pubvolume{1}
\issuenum{1}
\articlenumber{0}
\pubyear{2020}
\copyrightyear{2020}
\history{Received: date; Accepted: date; Published: date}

\Title{Effect of Oxygen Mole Fraction on Static Properties of Pressure-Sensitive Paint}

\Author{Tomohiro Okudera $^{1}$, Takayuki Nagata $^{2}$\orcidN{}, Miku Kasai $^{3}$\orcidK{}, Yuji Saito $^{4}$\orcidS{}, Taku Nonomura $^{5}$\orcidT{}, Keisuke Asai $^{6}${}}

\AuthorNames{Tomohiro Okudera, Takayuki Nagata, Miku Kasai, Yuji Saito, Taku Nonomura, and Keisuke Asai}

\address{%
$^{1}$ \quad Department of Aerospace Engineering, Tohoku University; okudera.tomohiro@aero.mech.tohoku.ac.jp (T.O.)\\
$^{2}$ \quad Department of Aerospace Engineering, Tohoku University; nagata@tohoku.ac.jp (T.Na.)\\
$^{3}$ \quad Department of Aerospace Engineering, Tohoku University; kasai.miku@aero.mech.tohoku.ac.jp (M.K.)\\
$^{4}$ \quad Department of Aerospace Engineering, Tohoku University; yuji.saito@tohoku.ac.jp (Y.S.)\\
$^{5}$ \quad Department of Aerospace Engineering, Tohoku University; nonomura@tohoku.ac.jp (T.No.)\\
$^{6}$ \quad Department of Aerospace Engineering, Tohoku University; keisuke.asai.c3@tohoku.ac.jp (K.A.)}

\corres{Correspondence: okudera.tomohiro@aero.mech.tohoku.ac.jp; Tel.:+81-22-795-4075}

\abstract{The effects of oxygen mole fraction on the static properties of pressure-sensitive paint (PSP) were investigated. Sample coupon tests using a calibration chamber were conducted for polymer-based PSP (PHFIPM-PSP), polymer/ceramic PSP (PC-PSP), and anodized-aluminium PSP (AA-PSP). The oxygen mole fraction was set to be between 0.1--100\% and the ambient pressure was set to be between 0.5--140~kPa. The localized Stern--Volmer coefficient $B_{\rm local}$ once increases and then decreases as the oxygen mole fraction increases. The value of $B_{\rm local}$ depends on both ambient pressure and oxygen mole fraction, but the effect of this parameter can be characterized as a function of the partial pressure of oxygen. The value of $B_{\rm local}$ of AA-PSP and PHFIPM-PSP, which are low-pressure type and relatively low-pressure type PSP, have a peak at the relatively low partial pressure of oxygen, and $B_{\rm local}$ of PC-PSP, which are atmospheric pressure type PSP, has a peak at the relatively high partial pressure of oxygen. The peak of the intensity change with respect to pressure fluctuation proportional to the ambient pressure $S_{\mathcal{PR}}$ appears at the lower partial pressure of oxygen than that of $B_{\rm local}$. This is because the intensity of PSP becomes quite low at the high partial pressure of oxygen even if $B_{\rm local}$ is higher. Hence, an optimal partial oxygen mole fraction exists depending on the type of PSP and ambient pressure range of the experiment, and its optimal value can be found based on the partial pressure of oxygen.}

\keyword{pressure-sensitive paint; oxygen mole fraction; optimal condition}  

\begin{document}


\section{Introduction}
Measurements of state quantities are one of the most important issues in scientific research and engineering fields. The measurement of distributions of state quantities such as pressure, density, and velocity in the fluid dynamic field is important for the performance evaluation of a fluid machine and clarifying flow dynamics. Therefore, aerodynamic measurement techniques are an essential tool. One of the most important measurement targets is surface pressure distributions, but complex models, which have many pressure taps and tubes, are required for the acquisition of pressure distribution. In the case of small or thin test models, it is extremely difficult to carry out multi-point measurements.

Pressure-sensitive paint (PSP)~\cite{liu1998luminescent} is an effective tool for measuring surface pressure distributions with high spatial resolution without complicated structure for pressure tubes in wind tunnel models. Pressure measurements using PSP are achieved by measuring the intensity or lifetime of PSP emissions under a constant oxygen mole fraction. The composition of PSP is oxygen-sensitive dye molecules, which are quenched by interaction with oxygen molecules, and a binder to fix the dye molecules to model surfaces. The characteristics of PSP can be changed depending on the combination of dye molecules, binder, any additives, and other such components.

Aerodynamic measurements and flow visualization using PSP have been conducted to various measurement targets and conditions~\cite{westerweel2004single,merienne2015pressure,peng2016fast,nakakita2002pressure,sugioka2019flight,masini2020analysis,huang2015applications}.
Of the various parameters of the test condition, the range of ambient pressure is one of the critical parameters on PSP measurements because it has a large impact on the characteristics of PSP such as the pressure sensitivity, intensity of emission, and so on.
In the case of low-pressure conditions, Anyoji et al.~\cite{anyoji2015pressure,anyoji2014effects} conducted PSP measurements on a flat plate under compressible low-Reynolds-number conditions. They acquired a time-averaged pressure distribution at the ambient pressure between 2 and 20~kPa. However, time-resolved PSP measurements in low-pressure and subsonic conditions are further difficult. Nagata et al.~\cite{nagata2020experimental} conducted PSP measurements on the surface of a circular cylinder at several kPa. They successfully extracted the characteristic modes of pressure fluctuation from a large number of time-series low signal-to-noise ratio (SNR) PSP images by applying randomized singular value decomposition~\cite{halko2011finding}. Niimi et al.~\cite{niimi2005application} applied PSP for measurements in the high-Knudsen number regime, which is rarefied gas flow mainly lower than 150~Pa. Also, Mitsuo~\cite{mitsuo2014characteristics} investigated the characteristics of PSP in the ultra low-pressure range of $10^{-3}$--$10^2$~Pa.

Aerodynamic measurements on high-Reynolds number flows or real-flight Reynolds number are conducted in high pressure ($>100$~kPa) to increase the Reynolds number~\cite{sugioka2018polymer}. One of the critical issues on PSP measurements in high pressure is a decrease in the intensity of PSP emission. In the case of a cryogenic wind tunnel, the low-pressure type PSP, such as AA-PSP and PTMSP-PSP, is used even though the wind tunnel working gas is pressurized~\cite{sakaue2001open,asai2002novel,egami2006appropriate}. This is because the oxygen mole fraction of the wind tunnel working gas is quite low (approximately 0.1\%). The characteristics of PSP change due to oxygen mole fraction even if the ambient pressure is the same. The characteristics of PSP under lower oxygen mole fractions have been investigated and applied. Asai et al. and Sakaue et al.~\cite{asai1997surface,sakaue2001open} developed AA-PSP for a cryogenic wind tunnel. They investigated the characteristics of various kinds of AA-PSP by sample coupon tests and clarified the influence of oxygen mole fraction on the pressure sensitivity at 100~K in the oxygen mole fraction ranging from 4~ppm to 2000~ppm and stagnation pressure ranging from 0.4 to 190~kPa.
Asai et al.~\cite{asai2002novel} developed (poly)TMSP-based PSP for cryogenic and unsteady tests. They conducted sample coupon tests at cryogenic temperatures and low oxygen mole fraction of less than 0.1\% and applied developed PSP into wind tunnel experiments. Egami, Fey, and Engler~\cite{egami2006appropriate} investigated the appropriate combination of dye molecules and a binder for PSP measurement under cryogenic temperature and low oxygen mole fraction of less than 0.1\%. Ono et al.~\cite{ono2010development} conducted sample coupon tests in the mixture of CO$_2$ and O$_2$ with the oxygen mole fraction of 0.1\% for PSP measurement in a wind tunnel simulating the atmospheric environment of Mars.

Oglesby, Puram, and Upchurch~\cite{oglesby1995optimization} derived the equations for relative error in pressure as a function of normalized intensity of PSP emission, the relative error in pressure as a function of pressure, and the relationship between sensitivity and pressure based on the Stern--Volmer equation. Their analysis provided fundamental limits to achievable sensitivity and accuracy in the considered model. Nagata et al.~\cite{nagata2020optimum} investigated an optimal pressure range of several types of PSP in the air using parameters proposed by them. They showed that the existence of the optimal pressure range from several points of view such as the Stern--Volmer coefficient and the intensity change in PSP emission against normalized pressure change. 

Since PSP is a sensor that utilizes oxygen quenching, the optimal pressure should correspond to the optimal partial pressure of oxygen (or oxygen mole fraction). There are several studies on the characteristics of PSP under the conditions of the oxygen mole fraction different from that of the air as mentioned above, but there is no research regarding the effect of oxygen mole fraction on the characteristics of PSP and no systematic report regarding the effect of oxygen mole fraction in a wide range of conditions.
In the present study, the effect of oxygen mole fraction on the static properties of PSP is investigated. Sample coupon tests are conducted for polymer-based PSP (PHFIPM-PSP), polymer/ceramic PSP (PC-PSP), and anodized aluminium PSP (AA-PSP) under several oxygen mole fractions in a wide range of ambient pressure. The static properties are evaluated based on the Stern--Volmer coefficient commonly used in the wind-tunnel community $B$, the localized Stern--Volmer coefficient $B_{\rm local}$, and the intensity change with respect to pressure fluctuation proportional to the ambient pressure $S_{\mathcal{PR}}$. The aim of the present study is to provide a way to improve PSP measurements in severe conditions by tuning the oxygen mole fraction of the working gas.

\section{Principal of PSP and Evaluation Parameters}
\subsection{Principal of PSP}
Pressure sensitivity is based on oxygen quenching. It theoretically obeys the following Stern--Volmer equation:
\begin{equation}
\label{eq:SVeq}
\frac{I_0}{I(P,T)}=1+K(T)P,
\end{equation}

\begin{equation}
\label{eq:SVcoef}
K(T)=\phi_{\rm O_2}S(T)K_{\rm SV},
\end{equation}
where $I_0$ and $I$ are the intensities of PSP emissions at the oxygen-free condition and at the pressure $P$, respectively; $S(T)$ is the solubility coefficients, which depends on the temperature $T$ of the binder; and the coefficients $K(T)$ and $K_{\rm SV}$ are the Stern--Volmer coefficients to the ambient pressure and oxygen mole fraction $\phi_{\rm O_2}$, respectively. In common wind tunnel experiments, Eq.~(\ref{eq:SVeq}) is normalized by the reference condition, which is generally wind-off condition, $P_{\rm ref}$ and $I_{\rm ref}$ because it is difficult to produce conditions without oxygen. By normalization of Eq.~(\ref{eq:SVeq}) using the wind-off condition ($I_{\rm ref}$ and $P_{\rm ref}$), the following equation can be obtained:
\begin{equation}
\label{eq:SVrel}
\frac{I_{\rm ref}}{I(P,T)}=A(T)+B(T)\left(\frac{P}{P_{\rm ref}}\right).
\end{equation}

\begin{equation}
\label{eq:SVcoef2}
\left\{
\begin{aligned}
A(T)=\frac{1}{1+K(T)P_{\rm ref}}, \\
B(T)=\frac{K(T)P_{\rm ref}}{1+K(T)P_{\rm ref}}.
\end{aligned}
\right.
\end{equation}
where $A(T)$ and $B(T)$ are the Stern--Volmer coefficients commonly used in the wind-tunnel community. These coefficients have temperature-dependent caused by thermal quenching.

The behavior of PSP ideally described by the linear model of Eq.~(\ref{eq:SVrel}), but the characteristics of the practical PSP have nonlinearity. Therefore, Eq.~(\ref{eq:SVrel}) is $I_{\rm ref}/{I}=f(P/P_{\rm ref})$ in the practical PSP where $f$ is a nonlinear function. The following equation that is similar to Eq.~(\ref{eq:SVrel}) can be obtained by the first order Taylor series approximation,
\begin{equation}
\label{eq:SVrellocal}
\frac{I_{\rm ref}}{I}=A_{\rm local}+B_{\rm local}\frac{P}{P_{\rm ref}},
\end{equation}
where $A_{\rm local}$ and $B_{\rm local}$ are the localized Stern--Volmer coefficients. The characteristics of the tested PSP will be mainly discussed based on this equation and $B_{\rm local}$ and related parameters of $B_{\rm local}$ are used as evaluation parameters. The evaluation parameters are introduced in the following subsection. In addition, Eq.~(\ref{eq:SVrellocal}) (also Eq.~(\ref{eq:SVrel})) can be rewritten as a function of the pressure coefficient $C_P$, the Mach number $M$, and specific heat ratio $\gamma$ in the variable pressure condition of the wind tunnel experiments as follows \cite{nagata2020optimum}:
\begin{equation}
\label{eq:nondimSVrel}
\frac{I_{\rm ref}}{I}=A_{\rm local}+B_{\rm local}\left(1+\frac{\gamma}{2}M^2C_P\right).
\end{equation}
Here, $C_P$ is the pressure coefficient,
\begin{equation}
\label{eq:cp}
C_P=\frac{P-P_{\infty}}{1/2\rho_{\infty} u_{\infty}^2}=\frac{P-P_{\rm ref}}{1/2\rho_{\infty} u_{\infty}^2},
\end{equation}
where $\rho_{\infty}$ and $u_{\infty}$ are the density and velocity in the mainstream in the case of wind tunnel experiments, respectively.
This equation indicates that $I_{\rm ref}/I$ does not explicitly depend on the ambient pressure $P_{\rm ref}$, but on the Mach number and the $C_p$ distribution. It should be noted that order of $C_p$ is almost the constant, but its distribution is affected by the ambient pressure $P_{\rm ref}$ through the change in the Reynolds number. It should be noted that the slight change in the distribution of $C_p$ by a Reynolds number is one of interests in the variable pressure condition of wind tunnel experiments.

\subsection{Evaluation Parameters}
The characteristics of PSP are evaluated using two parameters introduced by Nagata et al.~\cite{nagata2020optimum}. These parameters evaluate the performance of PSP in terms of measuring pressure fluctuations due to fluid phenomena. The first parameter is $B_{\rm local}$, which corresponds to normalized sensitivity.
\begin{equation}
\label{eq:Blocal}
B_{\rm local}\equiv 
\left.\frac{\text{d}\{I(P_{\rm ref})/I(P)\}}{\text{d}(P/P_{\rm ref})}\right|_{P=P_{\rm ref}}
\end{equation}
This parameter allows us to investigate the ambient pressure dependence of the sensitivity of PSP regardless of the influence of ambient pressure on the magnitude of the pressure fluctuation caused by the fluid phenomena. The magnitude of the intensity change due to the pressure fluctuation is also important in the instantaneous measurement such as single-shot life-time-based measurements or time-resolved measurements. The second parameter $S_{\mathcal{PR}}$ is the intensity change with respect to pressure fluctuation proportional to the ambient pressure $\Delta P$. 
The definition of $S_{\mathcal{PR}}$ is following based on Eq.~(\ref{eq:SVrellocal}):
\begin{equation}
\label{eq:spr}
S_{{\mathcal{PR}}}\equiv-\left.\frac{\text{d} I}{\text{d}\mathcal{P}}\right|_{\Delta P=0}=\left.\frac{I_{\rm ref}B_{\rm local}}{\left(A_{\rm local}+B_{\rm local}{\mathcal P}\right)^2}\right|_{\Delta P=0}=I_{\rm ref}B_{\rm local},
\end{equation}
where $\mathcal{P}=(P_{\rm ref}+\Delta P)/P_{\rm ref}=P/P_{\rm ref}$, and thus, the pressure fluctuation is normalized by the ambient pressure as in $B_{\rm local}$. This parameter indicates the gradient of the $P/P_{\rm ref}$--$I$ curve.

The larger these parameters in the certain ambient pressure are, the more advantageous it is for measurement at that ambient pressure. The optimal ambient pressure in terms of the maximization of these evaluation parameters can be derived analytically in the case of the linear model of Eq.~(\ref{eq:SVrel}). The parameter $B_{\rm local}$ corresponds to $B$ in the linear model, and $\mathcal{S_{PR}}$ becomes simply the following form,
\begin{align}
S_{\mathcal{PR}}\equiv&\left.-\frac{\text{d} I}{\text{d}\mathcal{P}}\right|_{\Delta P=0}=I_{\rm ref}B=\frac{I_0KP_{\rm ref}}{(1+KP_{\rm ref})^2}. \label{eq:sprlin2}
\end{align}

Obviously, the optimal ambient pressure in terms of the maximization of $B$ is $P=\infty$ from Eq.~(\ref{eq:SVcoef2}). The optimal ambient pressure in terms of the maximization of $S_{\mathcal{PR}}$ can be obtained by seeking the pressure at which the derivative of $S_{\mathcal{PR}}$ becomes zero as follows using Eq.~(\ref{eq:sprlin2}):
\begin{equation}
0 = \frac{{\rm d}{S_{\mathcal{PR}}}}{{\rm d} P_{\rm ref}}=\frac{{\rm d}}{{\rm d}P_{\rm ref}}\left\{\frac{I_0KP_{\rm ref}}{\left(1+KP_{\rm ref}\right)^2}\right\} = \frac{I_0K\left(1-KP_{\rm ref}\right)}{\left(1+KP_{\rm ref}\right)^3},
\end{equation}
and thus, $S_{\mathcal{PR}}$ takes a maximum at $P_{\rm ref}=1/K$. In addition, Oglesby, Puram, and Upchurch~\cite{oglesby1995optimization} previously discussed the optimal pressure in terms of the minimization of the relative error in measured pressure $\epsilon=\frac{\Delta P_{\epsilon}}{P}$. This can be derived by seeking the pressure at which the derivative of $\epsilon$ becomes zero \cite{oglesby1995optimization}. Here, $\Delta P$ is the error in the measured pressure $P$ due to the uncertainty in the measured PSP emission $\Delta\left(I/I_0\right)$.  Based on the leading term of the Taylor expansion of Eq.~(\ref{eq:SVeq}) with respect to $P$, the small finite errors in pressure due to uncertainty in the measured PSP emission is 
\begin{align}
\Delta P_{\epsilon} \approx \frac{\left(1+KP\right)^2}{K}\Delta\left(I/I_0\right),
\end{align}
and the relative error in the measured pressure can be written as
\begin{align}
\epsilon =\frac{\Delta P_{\epsilon}}{P_{\rm ref}} \approx &\frac{\left(1+KP_{\rm ref}\right)^2}{KP_{\rm ref}}\Delta \left(I/I_0\right),
\end{align}
where $P\approx P_{\rm ref}$.
Therefore, the optimal ambient pressure in terms of the minimization of relative error in measured pressure can be obtained by following equation,
\begin{align}
0 =& \frac{{\rm d}{\epsilon}}{{\rm d} {P_{\rm ref}}}\approx \frac{{\rm d}}{{\rm d}{P_{\rm ref}}}\left\{\frac{\left(1+KP_{\rm ref}\right)^2}{KP_{\rm ref}}\Delta \left(I/I_0\right)\right\} =\frac{\left(K^2-\frac{1}{P_{\rm ref}^2}\right)}{K}\Delta \left(I/I_0\right),
\end{align}
and the optimal ambient pressure is $P_{\rm ref}=1/K$. The relative error in measured pressure fluctuation caused by the fluid phenomena is also takes a minimum at $P_{\rm ref}=1/K$, because the magnitude of the pressure fluctuation is proportional to the ambient pressure. The ambient pressure $P_{\rm ref}=1/K$ that the parameters in the linear model take a maximum and minimum are summarized in Table~\ref{tab:optcond} 

\begin{table}[H]
\centering
\begin{tabular}{lll}
\hline
Parameter & Maximum cond. & Minimum cond. \\ \hline
Stern-Volmer coefficient $B$ & $\infty$ & 0 \\
Intensity change with normalized pressure fluctuation $S_{\mathcal{PR}}$ & $1/K$ & $0,\infty$ \\
Pressure sensitivity $S_P$ [1/kPa] \cite{oglesby1995optimization} & 0 & $\infty$ \\
Relative error in measured pressure $\epsilon$  \cite{oglesby1995optimization} & $0,\infty$ & $1/K$ \\ \hline
\end{tabular}
\caption{Optimum ambient pressure for each parameter in linear model.}
\label{tab:optcond}
\end{table}

These optimal conditions are based on the linear model under the constant oxygen mole fraction. Therefore, the optimal pressure changes due to the change in the oxygen mole fraction and nonlinearity in the characteristics of PSP.

\section{Experimental Setup}
\begin{figure}[H]
\centering
\includegraphics[scale=0.3]{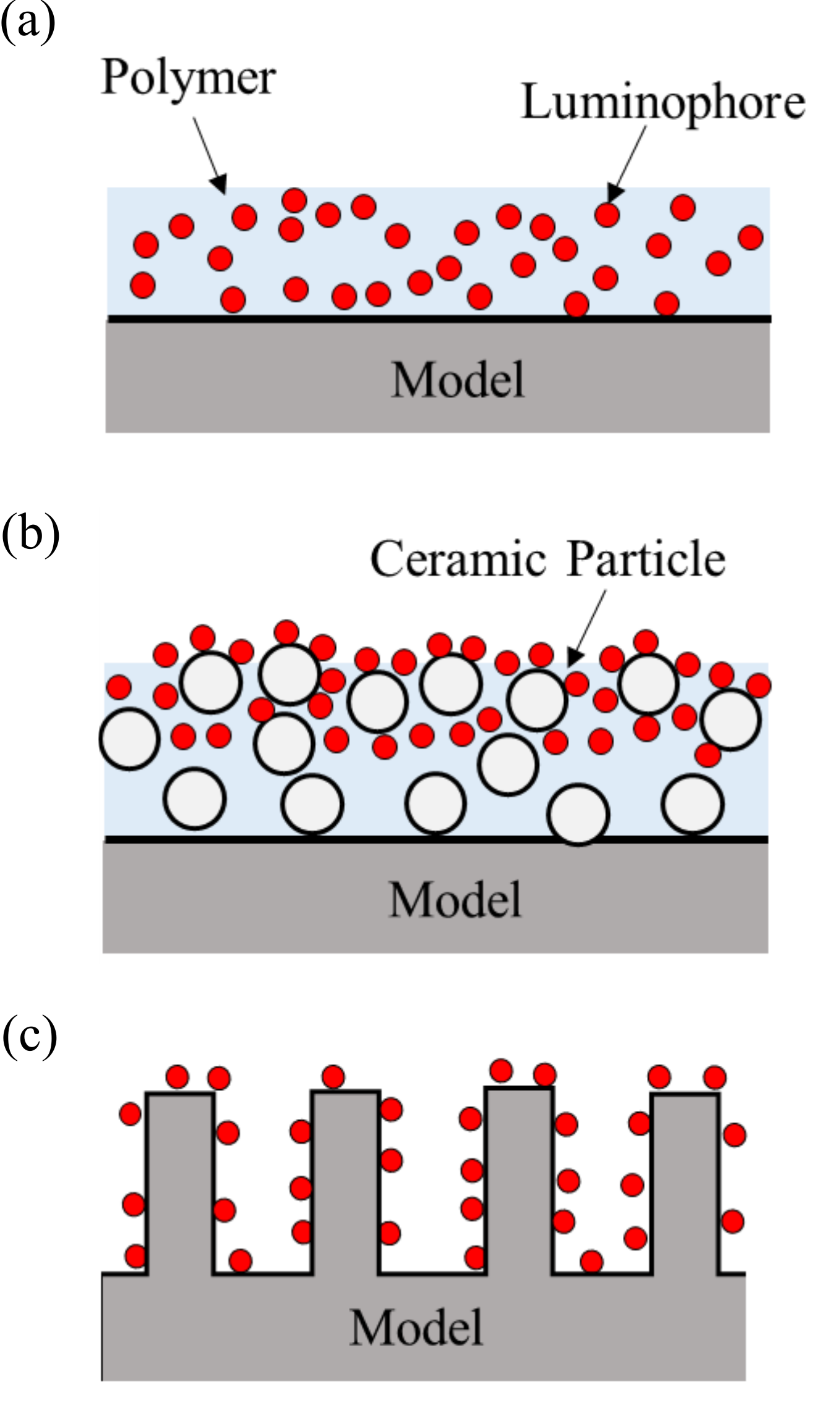}
\caption{Schematic diagrams of the binder structure of the tested PSP. (a) Polymer-based PSP (PHFIPM-PSP); (b) Polymer-ceramic PSP (PC-PSP); (c) Anodized-aluminium PSP (AA-PSP).}
\label{fig:binder}
\end{figure}

Sample coupon tests of three kinds of PSP, (poly)hexafluoroisopropyl methacrylate-based polymer PSP (PHFIPM-PSP)~\cite{mitsuo2014research}, PC-PSP~\cite{sugioka2016polymer} and AA-PSP~\cite{asai1997luminescent,sakaue2001time} were conducted in a calibration chamber. The compositions of the tested PSP are given in Table~\ref{tab:compo}. The schematic diagrams of the structure of the selected three kinds of PSP are shown in Figure~\ref{fig:binder}. The binder of PHFIPM-PSP is composed of polymer and dye molecules that are dispersing in the binder. This paint is a relatively low-pressure type and for steady measurements. Also, PC-PSP and AA-PSP, which are the atmospheric-pressure-type and the low-pressure-type PSP, are also tested. The binder of PC-PSP is a mixture of ceramic particles and polymer. The dye molecules exist relatively in the near-surface region. The binder of AA-PSP is anodized aluminum and the dye molecules exist on the surface of the binder.

\begin{table}[H]
\centering
\footnotesize 
\begin{tabular}{l|lll}
\hline
 & Polymer-based PSP (PHFIPM-PSP) \cite{mitsuo2014research} & PC-PSP \cite{sugioka2016polymer} & AA-PSP \cite{asai1997luminescent,sakaue2001time} \\ \hline
Binder & Poly(HFIPM) & \begin{tabular}[c]{@{}l@{}}Ester polymer\\ TiO$_2$ (rutile type)\end{tabular} & Anodized aluminum \\
Dye & PtTFPP & PtTFPP & Ru(dpp)$_3$ \\
Solvent & Ethyl acetate & \begin{tabular}[c]{@{}l@{}}Toluene (binder)\\ Toluene, Methanol (dye)\end{tabular} & Dichloromethane \\
Remarks & \begin{tabular}[c]{@{}l@{}}Low-temperature sensitivity\\ Relatively low-pressure type\end{tabular} & \begin{tabular}[c]{@{}l@{}}Fast responding\\ Atmospheric-pressure type\end{tabular} & \begin{tabular}[c]{@{}l@{}}Fast responding\\ Low-pressure type\end{tabular} \\ \hline
\end{tabular}
\caption{Composition of tested PSP.}
\label{tab:compo}
\end{table}

\normalsize

The sample coupon tests were conducted under the six oxygen mole fraction conditions of 0.1\%, 1\%, 10\%, 21\%, 40\%, and 100\%. The pressure inside the calibration chamber was varied in the ranges of $P=0.5$ to 30~kPa and $P=60$ to 140~kPa while the temperature of the sample coupon was kept at 293~K. The sample coupons were prepared for each condition of oxygen mole fraction to reduce the influence of photodegradation as much as possible. Polymer-based PSP was applied on the surface of each sample coupon at the same time. Each sample coupon of AA-PSP was created as a single large sample coupon until the anodizing process, and the absorption process of dye molecules was conducted for each separated sample coupon.

\begin{figure}[H]
\centering
\includegraphics[scale=0.5]{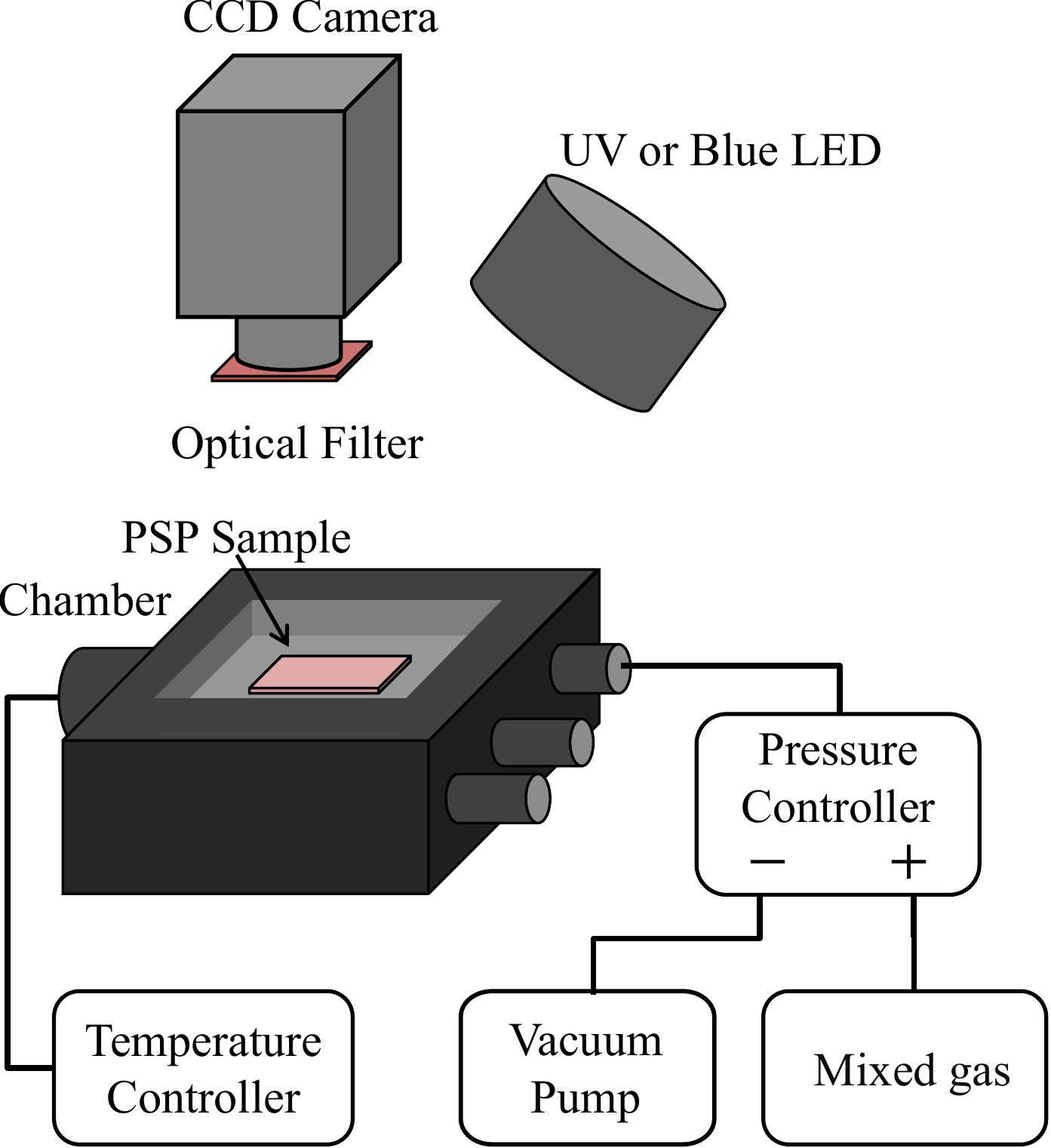}
\caption{Schematic diagram of the calibration chamber.}
\label{fig:cabchanber}
\end{figure}

The schematic diagram of the calibration chamber is shown in Figure~\ref{fig:cabchanber}. Pressure in the calibration chamber was controlled using the pressure controller (DPI 515, Druck), and the temperature of sample coupons was controlled by the temperature controller using the Peltier element (MT886-D1000, NetsuDenshi Kogyo). The precision of the pressure and temperature controllers is 30~Pa and 0.05~K, respectively. When the oxygen mole fraction was other than 21\%, the mixed gas cylinder of each mole fraction was used as the high-pressure source of the pressure controller. 
A 16-bit charge-coupled device (CCD) camera (ORCA II-BT1024, Hamamatsu Photonics) with a camera lens (Nikkor 105~mm f/2.8, Nikon) was used as a photodetector. For PHFIPM-PSP and PC-PSP, an ultraviolet LED (UV-LED; IL-106, Hardsoft) with a center wavelength of 395~nm was used as an excitation light source,  and an optical filter (650~nm $\pm$ 20~nm, PB0650/040, Asahi Spectra) was mounted in front of the camera lens. For AA-PSP, a blue light-emitting diode (LED) (LEDA294-470) with a center wavelength of 450~nm was used as an excitation light source, and an optical filter ($640\pm50$ nm, PB0640/100, Asahi Spectra) was mounted in front of the camera lens. The exposure time of the camera was set to achieve the PSP emission intensity of approximately 90\% of the full well capacity at the minimum ambient pressure for each oxygen mole fraction.

The air in the chamber was replaced for every pressure condition, and the oxygen mole fraction was precisely controlled. The mechanical shutter was implemented on the calibration chamber to reduce the influence of photodegradation as much as possible. The sample coupon was excited only at the moment of imaging. Moreover, the photodegradation correction was conducted by the linear interpolation in the time direction using the image acquired at the initial and end of the calibration tests for each oxygen mole fraction.

The uncertainty of $B_{\rm local}$ was calculated from the error propagation. The uncertainty of $B_{\rm local}$ stems from errors of the camera and pressure controller. The error of the camera includes dark current noise, readout noise and shot noise. Since the background is subtracted, the former two can be ignored. 
The uncertainty was approximately less than 5\%.

\section{Results and Discussion}
\begin{figure}[H]
\centering
\includegraphics[scale=0.6]{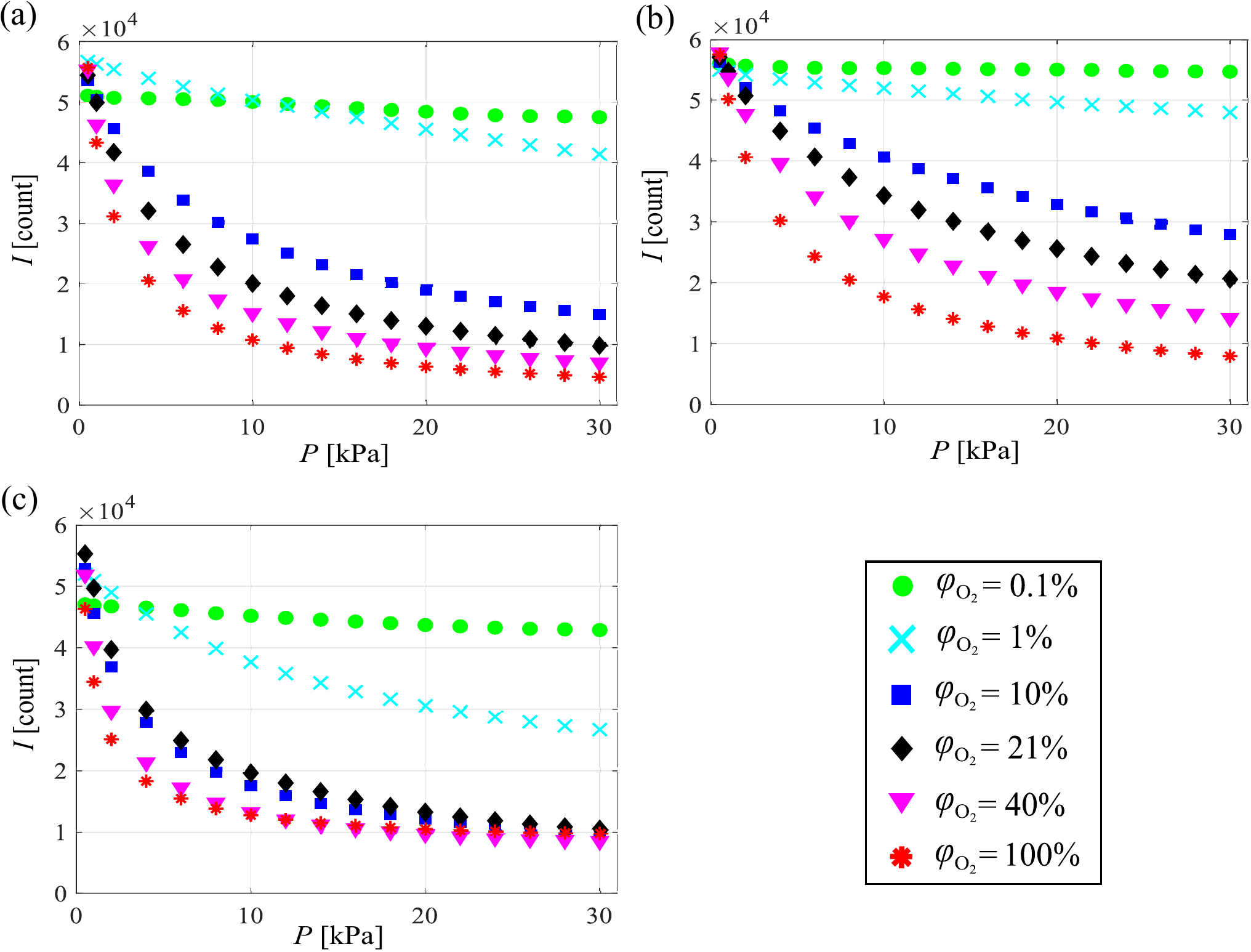}
\caption{Intensity of PSP emission for different oxygen mole fractions. (a)~PHFIPM-PSP; (b)~PC-PSP; (c)~AA-PSP.}
\label{fig:PI}
\end{figure}

Figure~\ref{fig:PI} shows the change in the emission intensity of three kinds of PSP, PHFIPM-PSP, PC-PSP, and AA-PSP, against the ambient pressure in the calibration chamber at $T=293$~K for different oxygen mole fractions. 

The PSP emission intensity decreases as the ambient pressure increases due to enhanced oxygen quenching. There is a large impact of the oxygen mole fraction on the gradient of the $P$--$I$ curves. The gradient of the curve becomes large and small at the high and low oxygen mole fractions, respectively, because the change in the partial pressure of oxygen corresponding to the number of oxygen molecules in the binder at a steady state depends on the oxygen mole fraction. The difference in the change rate of the emission intensity of PSP is related to the pressure sensitivity. Simultaneously, PSP measurements at high oxygen mole fractions are difficult due to the smaller emission intensity, especially for the measurements which have a limitation on exposure time such as time-resolved measurements.

\begin{figure}[H]
\centering
\includegraphics[scale=0.6]{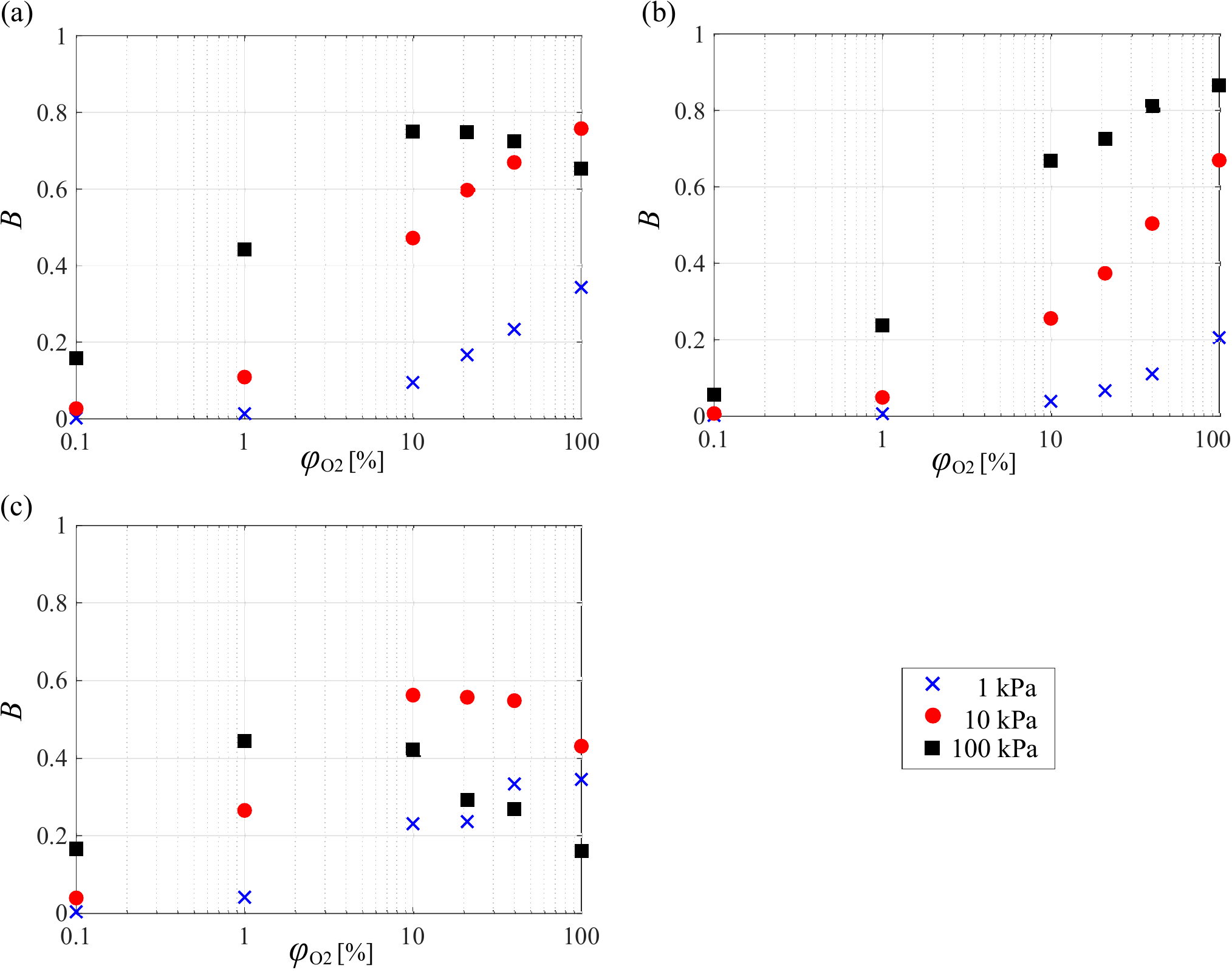}
\caption{Effect of oxygen mole fractions on $B$ at different pressure ranges. (a)~PHFIPM-PSP; (b)~PC-PSP; (c)~AA-PSP.}
\label{fig:B}
\end{figure}

Figure~\ref{fig:B} depicts the effects of the oxygen mole fraction on $B$ at different pressure ranges. Here, $B$ was computed by linear or quadratic fitting in the Stern--Volmer curves ($P/P_{\rm ref}$--$I_{\rm ref}/I$ curves) for each oxygen mole fraction. The linear fitting was used for PHFIPM-PSP and PC-PSP, and the quadratic fitting was used for AA-PSP. As a rough trend, $B$ increases as the oxygen mole fraction increases, and the baseline of $B$ becomes high at higher ambient pressure. However, the detailed trend is different for each PSP and ambient pressure range. For all PSP, $B$ at the ambient pressure of 1~kPa continuously increases as the oxygen mole fraction increases. This trend is the same at the ambient pressure of 10 and 100~kPa for PC-PSP and the ambient pressure of 10~kPa for PHFIPM-PSP. However, in other cases such as PHFIPM-PSP at 100~kPa and AA-PSP at 10 and 100~kPa, $B$ has a peak against the oxygen mole fraction for each ambient pressure. This result indicates that the PSP measurement of the fluid phenomena becomes better at appropriate oxygen mole fractions at certain ambient pressure in the tested atmospheric pressure range.

Changes in the oxygen mole fraction correspond to the change in the ambient pressure at constant oxygen mole fractions, in terms of the partial pressure of oxygen. Therefore, an optimal oxygen mole fraction (partial pressure of oxygen) should exist the same as the existence of the optimal ambient pressure in the air. Since the partial pressure of oxygen changes, the oxygen mole fractions that $B$ becomes a maximum varies also by a change in the ambient pressure.

The low and high oxygen mole fractions correspond to low and high ambient pressures for air, respectively. Hence, a decrease in $B$ by increasing oxygen mole fraction of AA-PSP, which has a higher $B$ at low ambient pressure in air, occurs despite low ambient pressure of 10~kPa. On the other hand, a decrease in $B$ of PC-PSP, which has a higher $B$ at around atmospheric pressure in air, does not occur in the investigated range of oxygen mole fraction and ambient pressure. However, $B$ of PC-PSP seems also to decrease with increasing the oxygen mole fraction in much higher ambient pressure. 

\begin{figure}[H]
\centering
\includegraphics[scale=0.6]{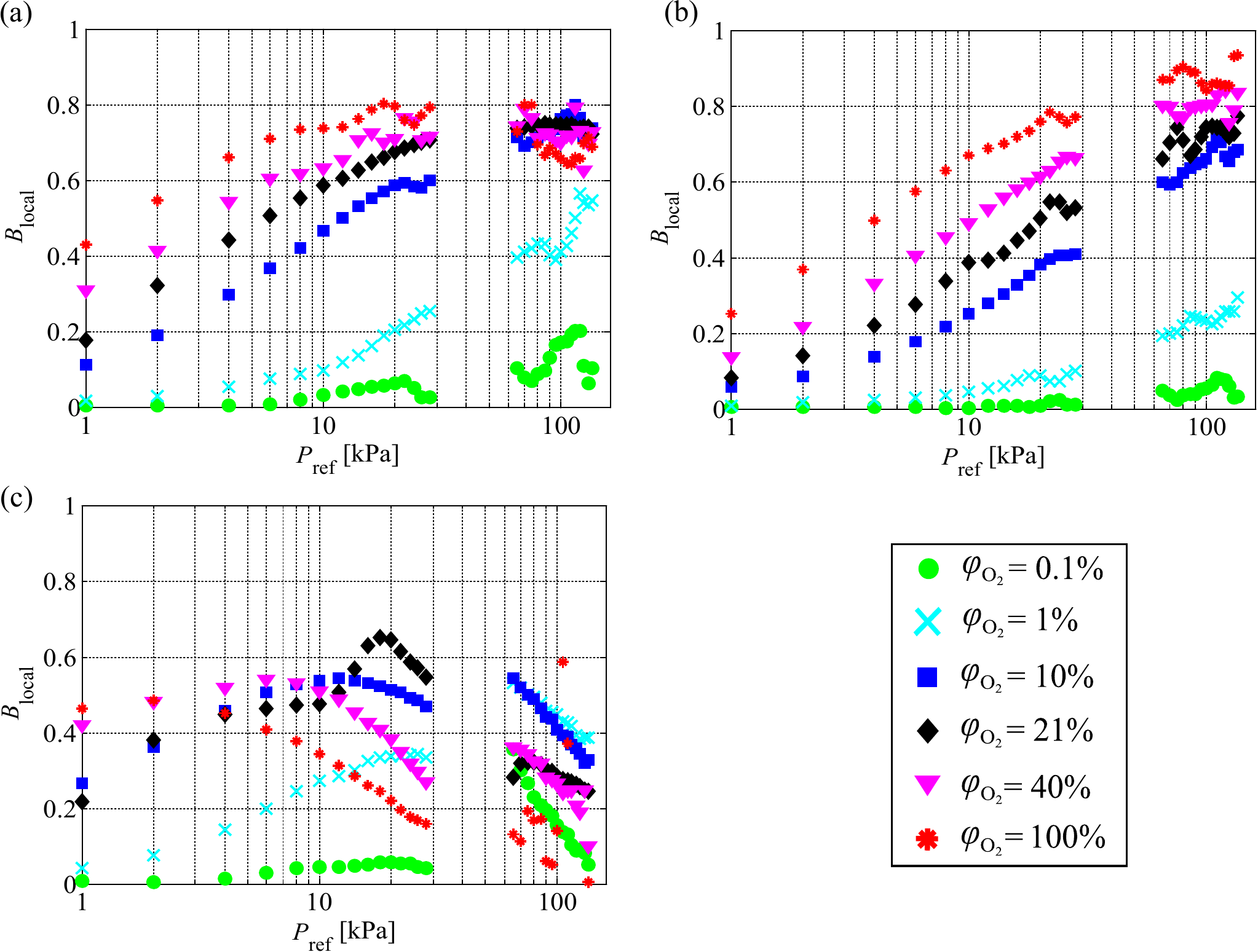}
\caption{Effect of the ambient pressure on $B_{\rm local}$ at different oxygen mole fractions. (a)~PHFIPM-PSP; (b)~PC-PSP; (c)~AA-PSP.}
\label{fig:Blocal_pt}
\end{figure}

\begin{figure}[H]
\centering
\includegraphics[scale=0.6]{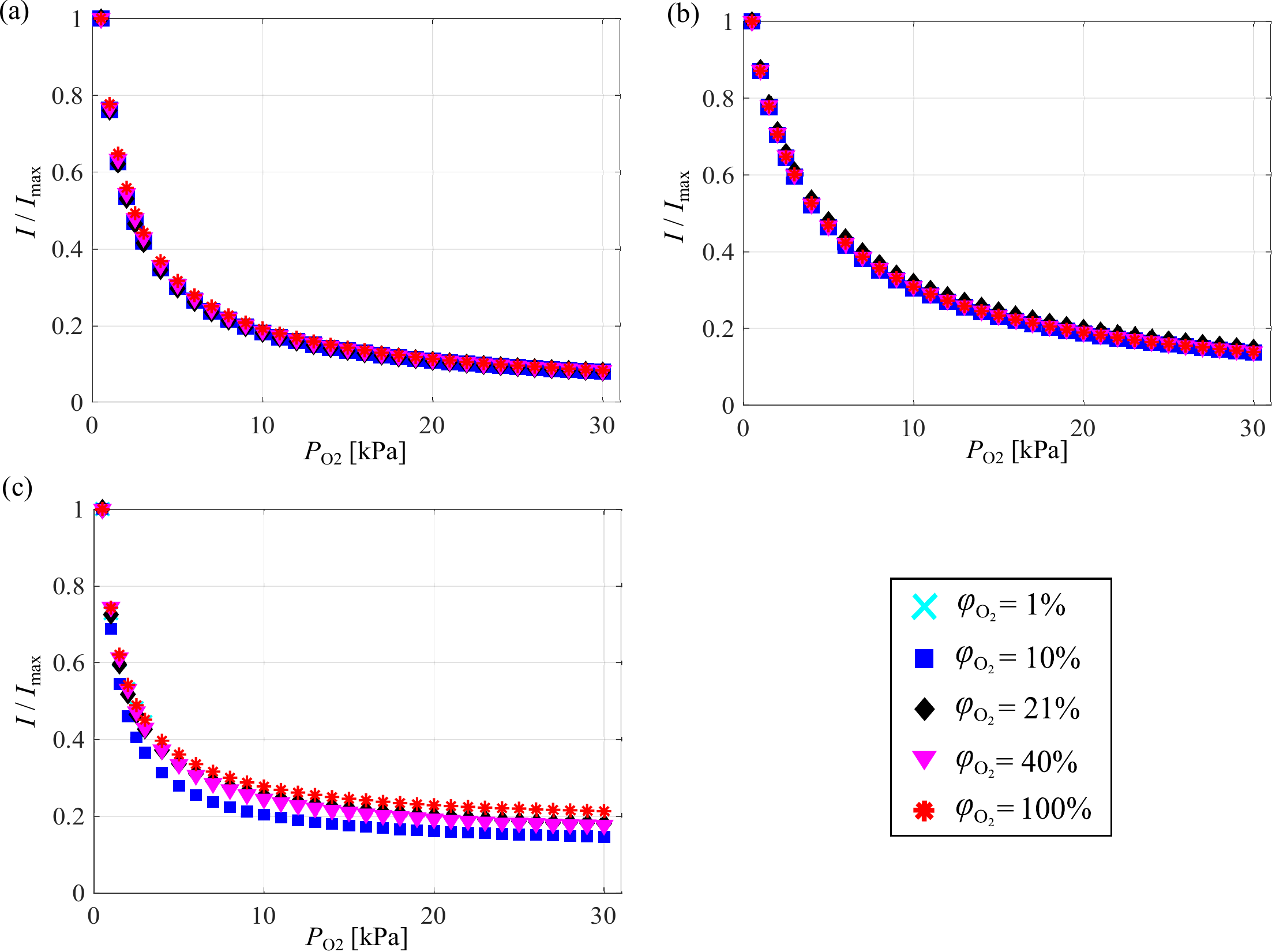}
\caption{Normalized intensity of PSP emission with respect to partial pressure of oxygen. (a)~PHFIPM-PSP; (b)~PC-PSP; (c)~AA-PSP.}
\label{fig:Po2I}
\end{figure}

Figure~\ref{fig:Blocal_pt} show the influence of the atmospheric pressure on $B_{\rm local}$ for several oxygen mole fractions. Focusing on PC-PSP, as a rough trend, $B_{\rm local}$ increases as the ambient pressure increases at each oxygen mole fraction. Also, $B_{\rm local}$ increases as the oxygen mole fraction increase at each ambient pressure, as discussed in Fig.~\ref{fig:B}. 
This is because the partial pressure of oxygen increases as the ambient pressure or oxygen mole fraction increases at each oxygen mole fraction. A similar trend as discussed in Fig~\ref{fig:B} can also be observed in the case of PHFIPM-PSP and AA-PSP.

%
%


Figure~\ref{fig:Po2I} shows the intensities of PSP emission, that are normalized by the maximum intensity in each oxygen mole fraction. The normalized intensity of PSP emission decreases as the partial pressure of oxygen decreases. Unlike in Fig.~\ref{fig:PI}, the normalized intensity at different oxygen mole fraction can be characterized by the partial pressure of oxygen.

\begin{figure}[H]
\centering
\includegraphics[scale=0.6]{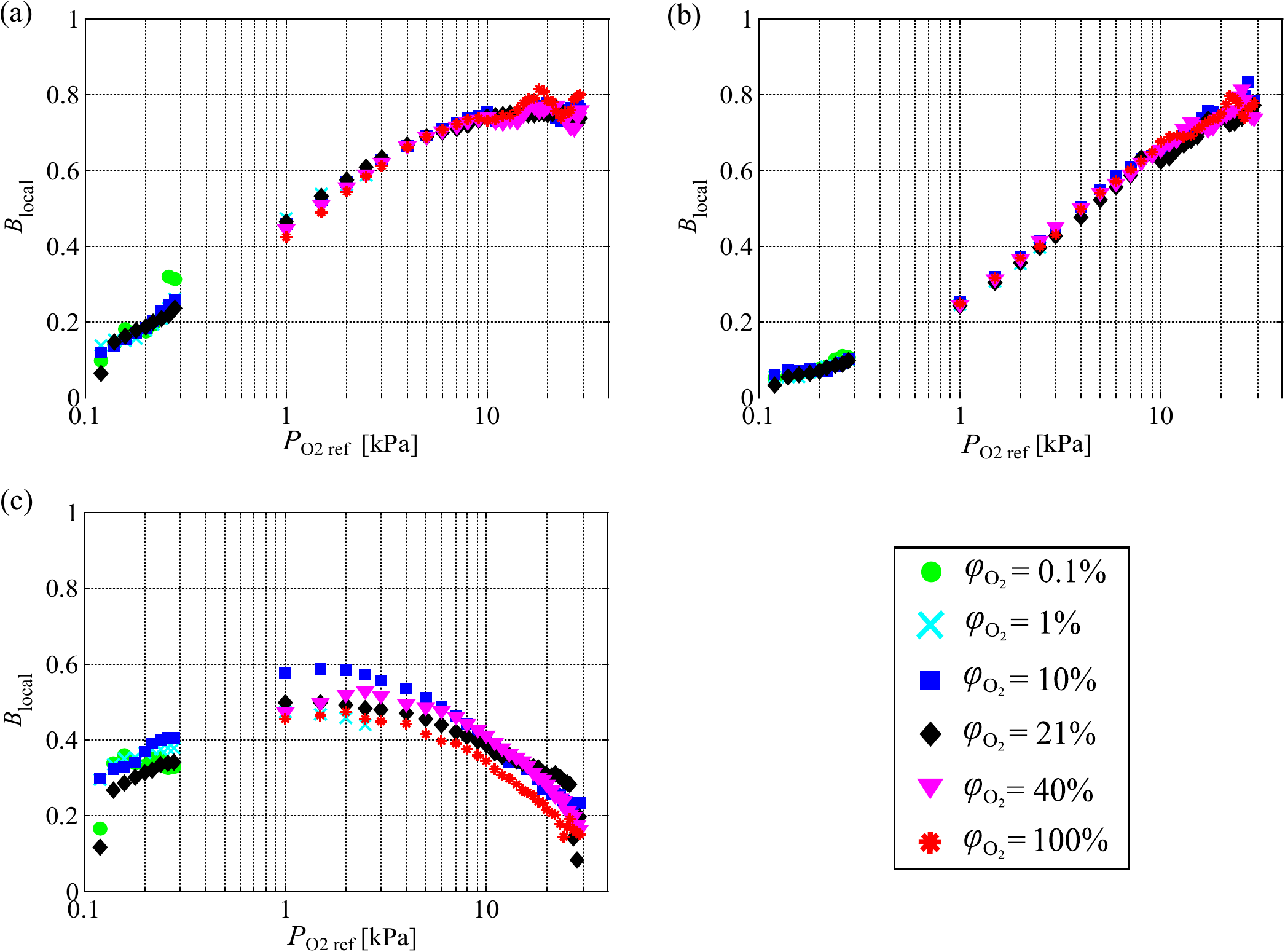}
\caption{Effect of the partial pressure of oxygen on $B_{\rm local}$ in different oxygen mole fraction. (a)~PHFIPM-PSP; (b)~PC-PSP; (c)~AA-PSP.}
\label{fig:B_Po2}
\end{figure}

Figure~\ref{fig:B_Po2} indicates the effect of the oxygen mole fraction on $B_{\rm local}$ as a function of the partial pressure of oxygen $P_{O_2}$. Curves of $B_{\rm local}$ for different oxygen mole fraction can be characterized by the partial pressure of oxygen. The values of $B_{\rm local}$ for all PSP increase as the oxygen mole fraction increase at the lower partial pressure of oxygen. In the tested condition, $B_{\rm local}$ has a peak, except PC-PSP. The partial pressure of oxygen which $B_{\rm local}$ becomes a maximum for PHFIPM-PSP was at approximately $P_{O_2}=20$~kPa. Such a peak appears in the much lower partial pressure of oxygen range in the case of AA-PSP. Although there is no peak in the range of the tested partial pressure of oxygen for PC-PSP, the peak can appear in the much higher partial pressure of oxygen.

The difference in the partial pressure of oxygen which $B_{\rm local}$ (and $B$) becomes a maximum reflects characteristics of PSP. In the case of PC-PSP, the relatively higher partial pressure of oxygen is required to dissolve oxygen molecules into the binder because the binder consists of Ester polymer and TiO$_2$ particles, and the solubility of oxygen of Ester polymer is low. On the other hand, the solubility of oxygen of poly(HFIPM), which is used in the binder of PHFIPM-PSP, is higher than the polymer used in PC-PSP, and thus oxygen can dissolve more at the same partial pressure of oxygen. When oxygen is sufficiently dissolved in the binder and becomes saturated, the aspect of oxygen quenching becomes difficult to change with respect to changes in the surrounding partial pressure of oxygen, and as a result, $B_{\rm local}$ decreases on the high partial pressure of oxygen side. This kind of behavior occurs at relatively low pressure or oxygen mole fraction when the solubility of oxygen is large. In the case of AA-PSP, dye molecules exist on the surface of the anodized aluminum binder. Dye molecules are exposed directly to the ambient gas in this case, and thus, decrease in $B_{\rm local}$ occurs even though partial pressure of oxygen is quite low. Hence, the optimal oxygen mole fraction exists depending on the type of PSP and ambient pressure range, and the optimal value can be characterized by the partial pressure of oxygen.

Here, the influences of the oxygen mole fraction observed by the experimental results are compared to that described by the linear model. According to the Smoluchowski relation, the Stern--Volmer coefficient to the pressure ($K$) of the polymer-based PSP is modeled as follows \cite{liu2005pressure}:
\begin{equation}
\label{eq:SVcoef3}
K=4\pi R_{\rm AB}N_0D\phi_{O_2}S,
\end{equation}
where $R_{\rm AB}$ is an interaction distance between dye molecules and oxygen molecules, $N_0$ is the Avogadro constant, and $D$ is the diffusivity. Based on Eq.~(\ref{eq:SVeq})--(\ref{eq:SVcoef2}) and \ref{eq:SVcoef3}, $B$ for the ideal polymer-based PSP can be characterized by the partial pressure of oxygen $P_{O_2}$, the solubility of oxygen $S$, and the diffusivity $D$, and $B$ increases as the partial pressure of oxygen, the solubility of oxygen, and diffusivity increases because of increasing $K$. This is related to the pressure which $B$ becomes a maximum is $P_{\rm ref}=\infty$ as shown in Table~\ref{tab:optcond}.
Such a trend qualitatively agrees with the present experimental result up to the partial pressure of oxygen which $B_{\rm local}$ becomes a maximum. However, $B_{\rm local}$ decreases at the higher partial pressure of oxygen because the oxygen molecule in the binder saturates. The partial pressure of oxygen which $B_{\rm local}$ coefficient becomes a maximum is considered to be depending on the solubility of oxygen and the diffusivity, i.e. oxygen permeability. For example, in the case of a binder with a large oxygen permeability, more oxygen molecules are taken into the binder and saturation occurs even with a small oxygen partial pressure. In this case, the peak of $B_{\rm local}$ appears at the lower partial pressure of oxygen.

The present result showed that the change in $B_{\rm local}$, including the effect of the saturation of the oxygen in the binder, can be described by the partial pressure of oxygen. The same explanation holds true for AA-PSP at the typical temperature which is described by the collision-controlled model. The oxygen permeability and diffusivity of the binder are not the parameters for the static properties of AA-PSP. However, the decreased partial pressure of oxygen which $B_{\rm local}$ becomes a maximum is considered to be related to the characteristics of the binder that the dye molecules exists on the surface of the binder. 

\begin{figure}[H]
\centering
\includegraphics[scale=0.6]{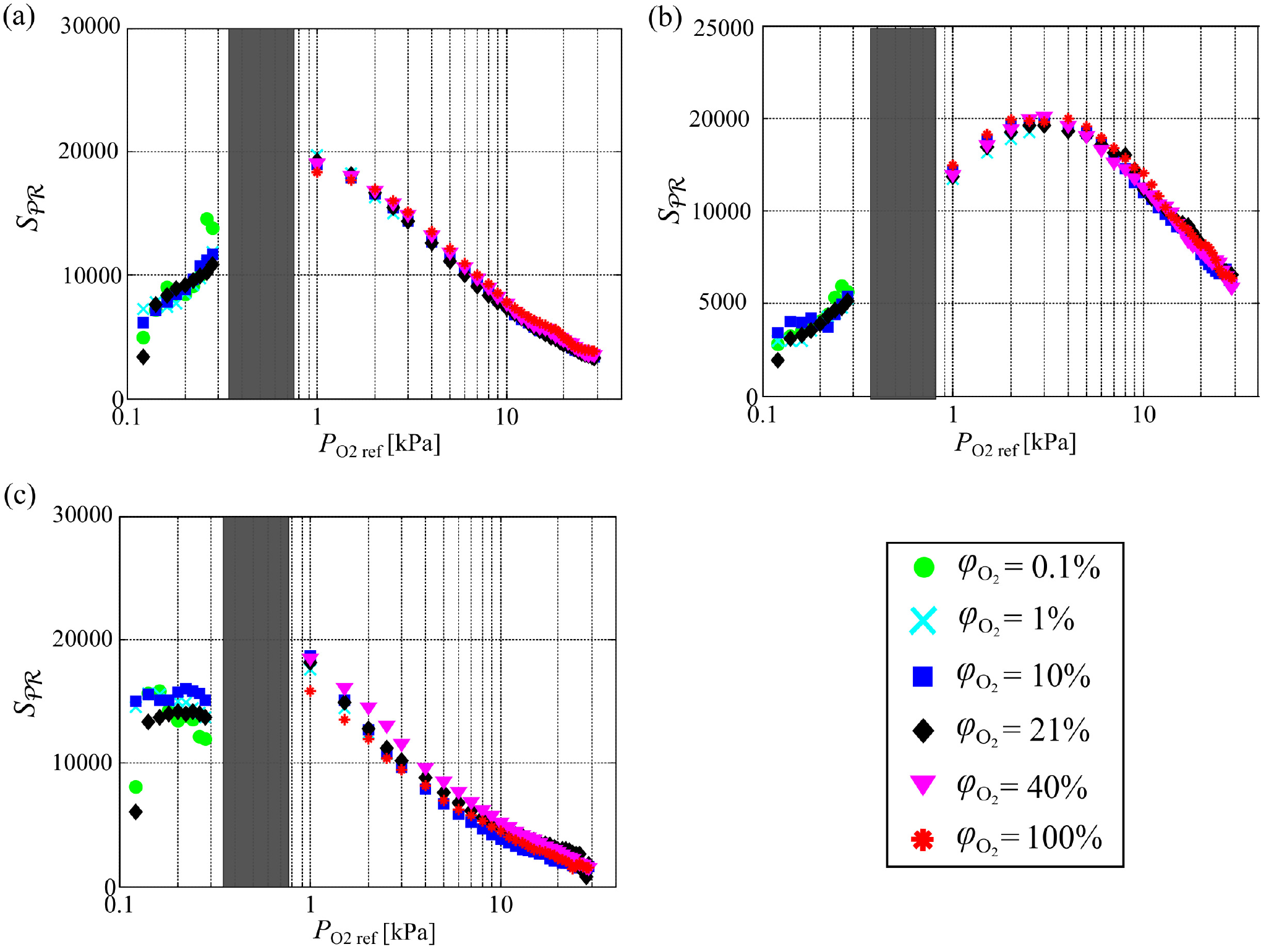}
\caption{Effect of the partial pressure of oxygen on $S_{\mathcal{PR}}$ at different oxygen mole fraction. (a)~PHFIPM-PSP; (b)~PC-PSP; (c)~AA-PSP.}
\label{fig:Spr_Po2}
\end{figure}

Figure~\ref{fig:Spr_Po2} indicates the effect of partial oxygen mole fraction on $S_{\mathcal{RP}}$. Here, $S_{\mathcal{RP}}$ includes the information regarding the intensity of PSP emission. The optical setup was changed for each PSP and between $P_{\rm{O}_2}<0.3$~kPa and $P_{\rm{O}_2}\geq1.0$~kPa, and thus the value of $S_{\mathcal{RP}}$ cannot compare between different PSP and the range of the partial pressure of oxygen between $P_{\rm{O}_2}<0.3$~kPa and $P_{\rm{O}_2}\geq1.0$~kPa.

By increasing the oxygen mole fraction, $B_{\rm local}$ increases up to certain partial pressure of oxygen, as discussed previously. However, the intensity of the PSP emission decreases as the partial pressure of oxygen increases. It is one of the critical issues as well as $B_{\rm local}$ because the smaller intensity of PSP emission causes a deterioration of SNR. Therefore, the intensity change against certain pressure changes is also an important property. The same as the curve of $B_{\rm local}$, the effect of oxygen mole fraction on $S_{\mathcal{RP}}$ can be characterized by the partial pressure of oxygen, and the partial pressure of oxygen that $S_{\mathcal{RP}}$ become a maximum is different for each PSP. Due to the nature of $S_{\mathcal{RP}}$, not only a larger $B_{\rm local}$ but also the larger intensity is required to increase $S_{\mathcal{RP}}$. Hence, the peaks of $S_{\mathcal{RP}}$ exist at the lower partial pressure of oxygen, and the peak can be confirmed including PC-PSP in the investigated range.

\section{Conclusions}
In the present study, we investigated the effect of oxygen mole fraction on the steady properties of three kinds of PSP by sample coupon tests. The oxygen mole fraction was set to be between 0.1--100\%, the ambient pressure was set to be between 0.5--140~kPa. The three kinds of PSP, PHFIPM-PSP, PC-PSP, and AA-PSP, were tested. The steady characteristics of PSP were evaluated by the Stern--Volmer coefficient $B$, the localized Stern--Volmer coefficient $B_{\rm local}$, and the intensity change with respect to pressure fluctuation proportional to the ambient pressure $S_{\mathcal{PR}}$.

The obtained data indicate that the intensity of PSP emission normalized by the maximum intensity in each oxygen mole fraction, can be characterized by the partial pressure of oxygen. By increasing the oxygen mole fraction, $B$ and $B_{\rm local}$ basically increases, but it becomes low again at further high oxygen mole fraction in the case of PHFIPM-PSP and AA-PSP. Hence, there is the optimal mole fraction, and the optimal oxygen mole fraction for each type of PSP depends on the ambient pressure. The Stern--Volmer curve can be characterized by the partial pressure of oxygen, and thus, the optimal condition can be found based on the partial pressure of oxygen regardless of ambient pressure. The partial pressure of oxygen that $B_{\rm local}$ becomes a maximum is considered to relate to the structure of the binder (e.g. oxygen permeability in the case of polymer-based PSP). Since the saturation of the oxygen in the binder occurs even though the partial pressure of oxygen is low, the partial pressure of oxygen that $B_{\rm local}$ takes a maximum is low when the oxygen permeability is high. In the case of AA-PSP, the partial pressure of oxygen that $B_{\rm local}$ becomes a maximum is further decreased than that of the polymer-based PSP. This is considered to be related to the characteristics of the binder that the dye molecules exist on the surface of the binder. The effect of oxygen mole fraction on $S_{\mathcal{PR}}$ was also investigated. This parameter involves both $B_{\rm local}$ and the intensity of PSP emission, and thus, the peak of the appears at the lower partial pressure of oxygen.

The localized Stern--Volmer coefficient $B_{\rm local}$ indicates the change in the normalized intensity of PSP emission against the change in the normalized pressure. Therefore, the optimal oxygen mole fraction can determine based on the relationship between $B_{\rm local}$ and the partial pressure of oxygen, when the intensity is not important such as the situation, which the on-tip integration of photon is valid. On the other hand, the magnitude of the intensity change due to the pressure fluctuation is important in the situation that the instantaneous measurement is required such as the single-shot life-time-based measurement or time-resolved measurement. In such a situation, the optimal oxygen mole fraction can be determined based on the relationship between the partial pressure of oxygen and $S_{\mathcal{PR}}$.

The present results show that the higher-oxygen mole fraction can provide better results in the case of the measurements using atmospheric pressure type PSP such as PC-PSP in low pressure. On the other hand, the lower-oxygen mole fraction is better for the measurements using low-pressure type PSP such as PHFIPM-PSP and AA-PSP in atmospheric or further high-pressure conditions.



\vspace{6pt} 



\authorcontributions{T.O., T.Na, and K.A. conceived and designed the experiments. T.O and M.K. performed the experiments under the guidance of K.A. and Y.S. T.O analyzed the data under the guidance of T.Na. K.A., T.No. T.O., T.Na, T.No, and K.A. contributed to the data interpretation. T.Na., T.O, and T.No contributed to writing the manuscript.}

\funding{The present work was supported by the Japan Society for the Promotion of Science, KAKENHI Grants 18K18906 and 19H00800.}

\reftitle{References}
\bibliography{xaerolab}
\end{document}